# IKNet: Interpretable Stock Price Prediction via Keyword-Guided Integration of News and Technical Indicators

Jinwoong Kim
Graduate School of Industrial Data Engineering
Hanyang University
Seoul, Republic of Korea
dnddl9456@hanyang.ac.kr

Sangjin Park*
Graduate School of Industrial Data Engineering
Hanyang University
Seoul, Republic of Korea
psj3493@hanyang.ac.kr

## Abstract

The increasing influence of unstructured external information, such as news articles, on stock prices has attracted growing attention in financial markets. Despite recent advances, most existing news-based forecasting models represent all articles using sentiment scores or average embeddings that capture the general tone but fail to provide quantitative, context-aware explanations of the impacts of public sentiment on predictions. To address this limitation, we propose an interpretable keyword-guided network (IKNet), which is an explainable forecasting framework that models the semantic association between individual news keywords and stock price movements. The IKNet identifies salient keywords via FinBERT-based contextual analysis, processes each embedding through a separate nonlinear projection layer, and integrates their representations with the time-series data of technical indicators to forecast next-day closing prices. By applying Shapley Additive Explanations the model generates quantifiable and interpretable attributions for the contribution of each keyword to predictions. Empirical evaluations of S&P 500 data from 2015 to 2024 demonstrate that IKNet outperforms baselines, including recurrent neural networks and transformer models, reducing RMSE by up to 32.9% and improving cumulative returns by 18.5%. Moreover, IKNet enhances transparency by offering contextualized explanations of volatility events driven by public sentiment.

*Corresponding author.

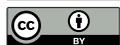

## 1 INTRODUCTION

Stock price predictions play a critical role in maximizing returns, optimizing asset allocation, and managing financial risk. However, markets are inherently complex and influenced by a wide range of factors, including geopolitical events, macroeconomic conditions, and investor sentiment, which introduces structural uncertainty [18, 27].

Among these factors, unstructured signals, such as political statements or breaking news, can trigger abrupt market swings. For instance, on April 2, 2025, U.S. President Trump announced a sweeping tariff policy that imposed a 10% base tariff on all imports and tariffs of up to 125% on Chinese goods. The S&P 500 index declined by 4.8% the following day, marking its sharpest single-day loss in 2020.[1] This example highlights the limitations of relying solely on structured data and underscores the importance of incorporating unstructured information into forecasting models.

Early forecasting methods relied on technical indicators, such as moving averages, the RSI, and the MACD, along with linear models, including ARIMA, GARCH, and ridge regression [2, 5, 15]. Although interpretable, these models assume linearity and struggle to capture nonlinear dynamics. Machine learning methods such as SVM, random forest, and XGBoost address some of these limitations by modeling complex relationships [7, 9, 17] but remain inadequate for capturing temporal dependencies inherent in financial time series data.

Recent advances in deep learning, such as recurrent neural networks (RNNs), temporal convolutional networks (TCNs), and transformers, have demonstrated strong performance in modeling long-range dependencies and have often outperformed traditional methods [4, 14, 19, 30]. However, these architectures are generally optimized for structured numerical inputs and struggle to effectively incorporate unstructured financial texts.

To bridge this gap, finance-specific language models, such as FinBERT [13, 31] have been used to extract sentiment-based features from financial news. These embeddings enable the integration of unstructured content into predictive pipelines. Although such models improve responsiveness to external signals, most rely on document-level sentiment scores or averaged embeddings that obscure the contributions of individual terms and hinder interpretability [23]. Previous studies demonstrate that incorporating financial news can improve forecasting accuracy and practical applicability [23, 32].

To overcome the interpretability limitations of prior approaches, we propose the interpretable keyword-guided network (IKNet),

---
[1]https://www.investopedia.com/dow-jones-today-04032025-11708250

which is a novel forecasting framework that integrates keyword-level features extracted from financial news with structured technical indicators to enhance both predictive performance and interpretability. The IKNet comprises three core components. First, FinBERT is used to identify and embed contextually relevant keywords from daily news articles. Second, each keyword embedding is passed through its own lightweight nonlinear projection layer, typically implemented as a small multilayer perceptron (MLP), which preserves separability and enables individual attribution while enhancing computational efficiency. Third, Shapley additive explanations (SHAP) [21] are applied to quantify and visualize the influence of each keyword on the prediction to thereby provide interpretable insights into how public sentiment impacts stock movements.

The main contributions of this study are as follows:

- We proposed IKNet, which is a forecasting framework that combines strong predictive accuracy with fine-grained interpretability via keyword-level attribution.
- We empirically evaluated the model using S&P 500 index data from 2015 to 2024, covering diverse market phases. IKNet consistently outperformed the baseline models in terms of both cumulative returns and predictive stability under volatile conditions.
- We employ SHAP-based visualizations to provide context-aware interpretations of model outputs to enable a transparent analysis of sentiment-driven volatility events.

The remainder of this paper is organized as follows. Section 2 reviews the related work. Section 3 introduces our dataset and its features. Section 4 describes the IKNet architecture. Section 5 presents experimental results. Section 6 discusses the key findings. Section 7 concludes the paper and discusses future directions.

## 2 RELATED WORK

### 2.1 Trading Systems in Stock Markets

Traditional trading systems generate signals based on historical data, such as asset prices and trading volumes [1, 22, 27]. Rule-based strategies that employ technical indicators, such as the MACD, RSI, ROC, and stochastic oscillators, are widely used to detect overbought or oversold conditions and have contributed to the rise of automated algorithmic trading [1, 22].

Although these approaches offer interpretability and modest predictive power, they struggle to capture abrupt market responses triggered by unstructured events such as news or policy announcements [18, 27]. To address this, recent studies have proposed machine learning- and deep learning-based models capable of learning complex nonlinear patterns from diverse financial and macroeconomic variables [19, 27]. Moreover, integrating external information with technical indicators has been shown to enhance forecasting accuracy [13].

### 2.2 Sentiment-based Stock Forecasting

Early forecasting models were predominantly statistical and linear, such as the ARIMA, GARCH, and ridge regression methods [2, 5, 15]. Although such models are interpretable and computationally efficient, they assume linearity and are inadequate for capturing the complexity of financial markets.

To overcome these limitations, machine learning methods, including artificial neural networks (ANNs), support vector machines (SVMs), random forests, and XGBoost were introduced [7, 9, 17, 27]. These techniques were followed by deep learning approaches, including RNNs, TCNs, and transformers that effectively model temporal dependencies from historical data [4, 14, 19, 30]. However, these models are typically restricted to structured inputs and have difficulty accounting for sudden market shifts driven by unstructured information, such as news or policy shocks [18, 23].

To address this issue, recent studies have incorporated sentiment analysis by leveraging domain-specific language models, such as FinBERT, to extract sentiment features from financial news and temporally align them with market data [13, 31]. Most approaches rely on document-level sentiment scores [23] or average embeddings [32], which have been shown to improve forecasting performance when combined with technical indicators [13, 32].

### 2.3 Explainable Artificial Intelligence in Stock Forecasting

Explainable artificial intelligence (XAI) methods are increasingly being applied to stock prediction models that incorporate sentiment features. Interpretability is especially critical in financial contexts, in which trust and transparency are essential for decision-making. Among widely used methods, such as SHAP and LIME [21, 29], SHAP has gained prominence owing to its game-theoretic foundation and consistent and intuitive visualization of feature contributions [21, 25].

Several studies used SHAP to analyze the impacts of sentiment scores, embeddings, and time series features on model predictions [26, 32]. However, most of these applications provide only high-level correlational insights and lack the fine-grained attribution of specific events to market movements [18, 26]. These results limit their utility as decision support tools because investors require not only accurate forecasts but also transparent identification of the key drivers behind them [18, 26].

To overcome this limitation, the proposed IKNet introduces independent nonlinear projection layers for each keyword, which thereby enables word-level attribution via SHAP. This design facilitates fine-grained interpretability and offers a clearer attribution of predictive contributions compared with conventional aggregation methods, such as average embeddings or attention-based fusion.

## 3 Dataset

This study constructs a regression dataset for stock price predictions by integrating S&P 500 index data with the corresponding daily financial news. To capture short-term market responses to external information, we define the prediction target as the next-day closing price [18]. The dataset spans 2015-2024 and covers distinct market phases: a relatively stable period (2015–2017) [16], trade tensions and economic slowdown (2018–2019) [24], and high volatility driven by the COVID-19 pandemic and subsequent recovery cycles (2020–2024) [3]. All data are aligned with U.S. trading days to ensure temporal consistency between the financial and textual inputs.

Daily financial news articles were retrieved from Google News[2] using the keyword "S&P 500." Full texts were extracted via HTML

---
[2]https://news.google.com/

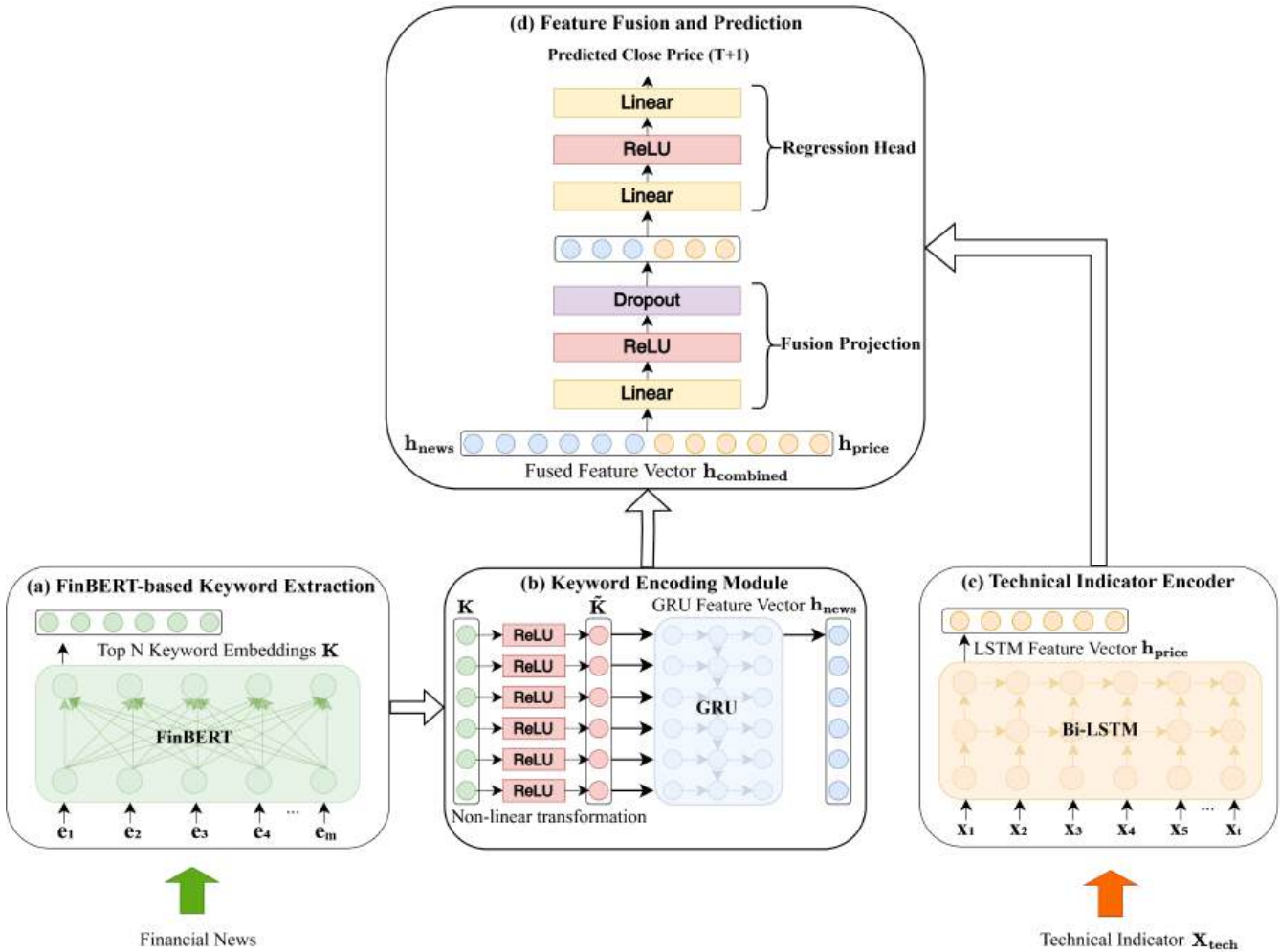

Figure 1: Architecture of the proposed IKNet model.

parsing, deduplicated, and filtered based on the presence of financial terms, such as "market", "economy", and "stock". After filtering, approximately 2,500 distinct articles were retained for analysis.

Technical indicators were derived from Yahoo Finance[3] using daily open, high, low, close, volume (OHLCV) data. Seventeen features were computed, including raw price and volume values, moving averages (SMA, EMA), momentum indicators (RSI, MACD, MACD Difference, signal line), volatility measures (Bollinger bands, Volatility ratio), volume change, and SMA-relative deviation.[4] [1, 22].

Each input sample comprises two components: (i) a multivariate time series of technical indicators and (ii) a set of keyword embeddings extracted from the news of the corresponding day. The output is the next-day closing price, which is treated as a continuous variable. This structure enables the model to capture structured and unstructured news-driven signals jointly.

For the evaluation, we adopted a sliding walk-forward validation strategy [12] that preserved temporal order and prevented data leakage. Each fold consisted of a three-year training window followed by a one-year test window, yielding a total of seven folds. For instance, Fold 1 trains in 2015-2017 and tests in 2018, Fold 2 shifts forward to train in 2016-2018 and test in 2019, and this pattern continues through 2024.

## 4 METHODOLOGY

This section describes the IKNet architecture, which is a regression model that predicts the next-day closing price by combining key keywords from financial news with technical indicators. As shown in Figure 1, IKNet comprises four modules: (a) FinBERT-based keyword extraction, (b) keyword encoding, (c) technical indicator encoding, and (d) feature fusion and prediction. The model processes unstructured text and structured time series data separately before integrating them. The arrows represent the sequential flow of information through each module.

---
[3] https://finance.yahoo.com
[4] Technical indicators such as SMA($n$=10), EMA($n$=10), RSI($n$=14), MACD(12,26,9), and Bollinger Bands($n$=20, $k$=2) were computed using the pandas-ta library.

## 4.1 FinBERT-based Keyword Extraction

To identify influential words in sentiment classification, we combined a pretrained FinBERT model [31] with a saliency-based approach. This method estimates the sensitivity of the model output with respect to each input token using gradient-based analysis, highlighting the tokens that contribute the most to the prediction. This method enhances interpretability by revealing the decision rationale of the model [20]. The saliency score for each token is calculated as follows:

$$s_i = \left\| \frac{\partial p}{\partial \mathbf{e}_i} \right\|_2, \quad (1)$$

where $s_i$ denotes the importance of the $i$-th token, $p$ is the FinBERT output, and $\mathbf{e}_i$ is the embedding vector of the $i$-th token. To quantify token importance, gradients were derived with respect to the logit of the predicted class, thereby excluding derivatives associated with non-predicted classes. The input sentences were first tokenized into sub-words and then merged into full words. Based on the average saliency scores, the top-$n$ tokens were selected as the most influential keywords. These keywords were passed through FinBERT again to obtain their embeddings, denoted as $\mathbf{K} = [\mathbf{k}_1, \mathbf{k}_2, \ldots, \mathbf{k}_n] \in \mathbb{R}^{n \times d}$, where $n$ is the number of keywords and $d$ is the embedding dimension. This module transforms unstructured text into structured vector representations for downstream encoding and prediction. The overall process is illustrated in module (a) of Figure 1.

## 4.2 Keyword Encoding Module

Each selected keyword embedding is independently passed through a nonlinear projection layer composed of a linear transformation, ReLU activation, and dropout. This process yields a lower-dimensional representation. Moreover, this design prevents representation mixing and enables the clear separation and quantitative assessment of word-level contributions to SHAP-based interpretations.

The projected embeddings are arranged sequentially and fed into a gated recurrent unit (GRU) that captures the implicit associations and sequential dependencies among keywords. The GRU, which is a variant of a recurrent neural networks (RNN), is well-suited for modeling time-dependent patterns. Compared with long short-term memory (LSTM), the GRU requires fewer parameters and offers greater computational efficiency, making it suitable for lightweight integration within multicomponent models [8].

In this study, the GRU aggregates information across the sequence and outputs a summary vector $\mathbf{h}_{\text{news}} \in \mathbb{R}^{2h}$. The overall process is as follows:

$$\tilde{\mathbf{k}}_i = \text{ReLU}(W \mathbf{k}_i + \mathbf{b}_0), \quad \mathbf{h}_{\text{news}} = \text{GRU}(\tilde{\mathbf{k}}_1, \ldots, \tilde{\mathbf{k}}_n), \quad (2)$$

where $\mathbf{k}_i \in \mathbb{R}^d$ is the embedding of the $i$-th keyword, $\tilde{\mathbf{k}}_i$ is its projected representation, $n$ is the number of selected keywords, $\mathbf{b}_0$ is the bias vector, and $h$ denotes the hidden dimension. This module corresponds to component (b) in Figure 1.

## 4.3 Technical Indicator Encoder

The input of technical indicators is represented as a time series of the past $T$ days, with the input matrix denoted as $\mathbf{X}_{\text{tech}} \in \mathbb{R}^{T \times f}$, where $f$ is the number of technical indicators used. In this study, we used the full set of price- and volume-based indicators defined earlier. This sequential input is processed using a bidirectional long short-term memory (Bi-LSTM) network. The Bi-LSTM is a type of recurrent neural network (RNN) that captures both past and future contexts, which makes it effective for modeling structured patterns over time. This property is particularly suitable for encoding technical indicators that exhibit continuous temporal variation [4, 14, 19].

At each time step, the forward and backward hidden states are concatenated, and an average pooling operation is applied over the entire sequence to generate a summary vector $\mathbf{h}_{\text{price}} \in \mathbb{R}^{2h}$. This process is formally defined as follows:

$$\overrightarrow{\mathbf{h}}_t = \text{LSTM}_{\text{fwd}}(\mathbf{x}_t), \quad \overleftarrow{\mathbf{h}}_t = \text{LSTM}_{\text{bwd}}(\mathbf{x}_t), \quad \mathbf{h}_{\text{price}} = \frac{1}{T} \sum_{t=1}^{T} [\overrightarrow{\mathbf{h}}_t; \overleftarrow{\mathbf{h}}_t], \quad (3)$$

where $\overrightarrow{\mathbf{h}}_t$ and $\overleftarrow{\mathbf{h}}_t$ denote the forward and backward hidden states of the LSTM at time step $t$, respectively. This process is illustrated in Figure 1 (c).

## 4.4 Feature Fusion and Prediction

The encoded representations from the news ($\mathbf{h}_{\text{news}}$) and technical indicators ($\mathbf{h}_{\text{price}}$) are concatenated to form a combined feature vector $\mathbf{h}_{\text{combined}} \in \mathbb{R}^{4h}$. This vector is passed through a nonlinear projection layer composed of a linear transformation, rectified linear unit (ReLU) activation, and dropout, which reduces its dimension to $\mathbb{R}^{2h}$. A regression head with two fully connected layers then produces the final predicted closing price $\hat{y} \in \mathbb{R}$. This process is formally expressed as follows:

$$\mathbf{h}_{\text{combined}} = [\mathbf{h}_{\text{news}}; \mathbf{h}_{\text{price}}], \quad \hat{y} = W_2 \cdot \text{ReLU}(W_1 \mathbf{h}_{\text{combined}} + \mathbf{b}_1) + \mathbf{b}_2, \quad (4)$$

where $W_1$ and $W_2$ are weight matrices, $\mathbf{b}_1$ and $\mathbf{b}_2$ are bias vectors, and ReLU is an activation function. This module captures the interactions between the heterogeneous features and generates the final prediction. The complete pipeline is shown in Figure 1 (d).

## 4.5 SHAP-based Analysis of Keyword Contributions

To interpret IKNet's predictions, we employ Kernel SHAP [21], a cooperative-game-theoretic XAI method that yields locally consistent feature attributions for arbitrary architectures. Specifically, it approximates the output function $f(\mathbf{z})$ of the model as a linear combination of the input features:

$$f(\mathbf{z}) = \phi_0 + \sum_{i=1}^{M} \phi_i z_i, \quad (5)$$

where $\mathbf{z} = [\mathbf{k}_1, \ldots, \mathbf{k}_n; \mathbf{x}_1, \ldots, \mathbf{x}_T]$ concatenates the top-$n$ keyword embeddings (Section 4.2) and the $T$-step technical-indicator vectors $\mathbf{x}_t \in \mathbb{R}^f$ (Section 4.3). Each scalar feature $z_i$ has an associated SHAP value $\phi_i$, and $\phi_0$ denotes the baseline output.

After training, we compute SHAP values on the test set. Mean absolute values quantify global feature importance, revealing the relative influence of keywords versus indicators. Token-level attributions are visualized per day; color and intensity respectively indicate contribution sign and magnitude, offering intuitive insight into how salient terms drive forecasts.

Table 1: Prediction performance based on the number of input keywords $K$ (RMSE / SMAPE) across test years.

| Year | $K$=9 | $K$=11 | $K$=13 | $K$=15 | $K$=17 | $K$=19 | $K$=21 |
|---|---|---|---|---|---|---|---|
| 2018 | **39.004** / 1.156 | 40.496 / 1.231 | 40.764 / 1.245 | 41.153 / 1.252 | 41.681 / 1.258 | 48.827 / 1.539 | 48.906 / 1.535 |
| 2019 | 37.628 / 1.086 | 36.770 / 1.079 | 34.949 / 1.023 | **32.007** / 0.906 | 32.398 / 0.924 | 38.901 / 1.142 | 48.475 / 1.457 |
| 2020 | 98.641 / 2.604 | 96.230 / 2.572 | 89.720 / 2.251 | 86.040 / 2.213 | 84.292 / 2.168 | 80.302 / 2.074 | **77.291** / 1.969 |
| 2021 | 152.125 / 3.001 | 135.438 / 2.684 | 126.049 / 2.467 | 107.777 / 2.204 | **74.923** / 1.531 | 176.455 / 3.684 | 180.686 / 3.896 |
| 2022 | 104.804 / 2.050 | 99.398 / 1.956 | 92.878 / 1.863 | 90.881 / 1.820 | 88.311 / 1.743 | **83.799** / 1.685 | 85.267 / 1.686 |
| 2023 | 49.660 / 0.929 | 50.122 / 0.922 | 49.233 / 0.924 | 50.339 / 0.938 | **48.135** / 0.906 | 50.268 / 0.947 | 53.359 / 1.002 |
| 2024 | 141.808 / 2.137 | 123.012 / 1.849 | 91.426 / 1.352 | 89.689 / 1.355 | **58.006** / 0.850 | 84.262 / 1.217 | 145.366 / 2.226 |

Table 2: Performance comparison by test year (RMSE / SMAPE) across baseline models and IKNet.

| Year | Ridge | LSTM | Transformer | TCN | FinBERT-Attention-LSTM | FinBERT-Sentiment-LSTM | IKNet (Ours) |
|---|---|---|---|---|---|---|---|
| 2018 | 57.780 / 1.791 | 49.634 / 1.478 | 46.360 / 1.254 | 44.302 / 1.282 | 46.631 / 1.445 | 43.718 / 1.215 | **41.681** / 1.258 |
| 2019 | 56.702 / 1.757 | 55.532 / 1.424 | 32.747 / 0.931 | 42.896 / 1.225 | 38.382 / 1.061 | 41.448 / 1.225 | **32.398** / 0.924 |
| 2020 | 114.230 / 2.967 | 87.379 / 2.113 | 81.245 / 1.993 | **71.167** / 1.644 | 101.198 / 2.436 | 86.542 / 1.903 | 84.292 / 2.168 |
| 2021 | 143.948 / 3.137 | 137.252 / 2.725 | 124.126 / 2.174 | 76.406 / 1.449 | 120.811 / 2.405 | 99.492 / 1.947 | **74.923** / 1.531 |
| 2022 | 111.539 / 2.283 | 93.109 / 1.863 | 107.601 / 2.141 | 97.022 / 1.913 | 105.808 / 2.149 | 106.553 / 2.167 | **88.311** / 1.743 |
| 2023 | 73.043 / 1.403 | 54.491 / 0.981 | 77.291 / 1.503 | **41.804** / 0.808 | 58.118 / 1.130 | 51.977 / 0.945 | 48.135 / 0.906 |
| 2024 | 150.386 / 2.474 | 127.877 / 2.092 | 142.280 / 2.113 | 85.559 / 1.311 | 118.971 / 1.679 | 87.575 / 1.196 | **58.006** / 0.850 |

## 5 EXPERIMENT

In this section, we quantitatively evaluate the predictive performance and interpretability of the proposed IKNet model. We conducted comparative experiments using the established time series models and hybrid models incorporating new features. We also assessed the robustness and practical utility of the model through a series of analyses, including performance variations across different keyword counts, ablation studies on keyword inputs, prediction visualizations, return-based simulations, and SHAP-based interpretability assessments.

### 5.1 Experimental Setup

The IKNet forecasts the next-day closing price using two types of inputs: (i) contextual embeddings of the top 17 keywords extracted from the financial news of the previous day and (ii) a set of technical indicators. A 10-day window was used to compute the indicators, which is consistent with standard practices for capturing short-term trends via metrics such as the simple moving average (SMA) and relative strength index (RSI) [1, 22].

Key hyperparameters were selected through a random search within a predefined space, and the same configuration was applied across all test years. The final architecture comprised one GRU layer and two Bi-LSTM layers, each of which had 256 hidden units and a dropout rate of 0.1. The model was trained using the Adam optimizer with a learning rate of 0.01, batch size of 32, and 200 epochs.

For comparison, six baseline models were implemented: (1) ridge regression [15], (2) LSTM [14], (3) transformer [30], (4) temporal convolutional network (TCN) [4], and two FinBERT-based models, (5) FinBERT attention LSTM [10], and (6) FinBERT sentiment LSTM [13]. Ridge regression is a linear model that uses L2 regularization to mitigate overfitting. The LSTM captures long-term dependencies in a time series, whereas the transformer models temporal relationships using parallel self-attention. A TCN employs causal and dilated convolutions for sequential pattern learning. Among these baselines, ridge regression, LSTM, transformer, and TCN rely solely on technical indicators as inputs, while the two FinBERT-based models additionally leverage financial news articles through a pretrained FinBERT model using sentiment scores or document-level embeddings.

In contrast, FinBERT-based baselines incorporate news information via a pretrained FinBERT model using sentiment scores or document-level embeddings. These serve as relevant benchmarks because they use input modalities similar to those of IKNet. All the baselines were implemented and trained in PyTorch using their original architectures under identical experimental settings.

### 5.2 Evaluation

The performance of IKNet was evaluated using the RMSE and SMAPE scores, with each experiment designed to quantitatively assess the predictive capability of the model in terms of input configuration and baseline comparison.

The prediction performances of IKNet with varying numbers of input keywords are summarized in Table 1. The number of keywords was adjusted from 9 to 21 in odd-numbered increments. For each setting, experiments were conducted based on the test samples from seven folds, covering the years 2018 (fold 1) to 2024 (fold 7). The results showed that using 17 keywords yielded the best overall performance, with an average RMSE of 61.107 and SMAPE of 1.340 across all years. These results suggest that an appropriate number of keywords can effectively summarize news information while suppressing irrelevant noise to thereby enhance predictive accuracy.

In general, the performance tends to degrade when the number of keywords is too small or large. The most stable results are obtained using a moderate number of keywords. Although year-specific fluctuations were observed, the highest average performance was

**Table 3: RMSE and SMAPE values of IKNet variants using Tech, Keyword, and Full inputs.**

| Year | Tech Only | Keyword Only | IKNet (Full) | DM Stat. (Tech) | DM Stat. (Keyword) |
|---|---|---|---|---|---|
| 2018 | 48.912 / 1.476 | 47.142 / 1.369 | 41.681 / 1.258 | -2.251* | -3.735* |
| 2019 | 54.838 / 1.564 | 53.036 / 1.529 | 32.398 / 0.924 | -5.818* | -9.323* |
| 2020 | 86.714 / 2.230 | 93.527 / 2.269 | 84.292 / 2.168 | -1.027 | -2.790* |
| 2021 | 134.823 / 2.755 | 113.220 / 2.266 | 74.923 / 1.531 | -6.370* | -5.983* |
| 2022 | 91.913 / 1.814 | 80.323 / 1.774 | 88.311 / 1.743 | -0.560 | 1.338 |
| 2023 | 53.964 / 1.016 | 48.897 / 0.916 | 48.135 / 0.906 | -1.624 | -0.229 |
| 2024 | 126.405 / 1.852 | 99.979 / 1.467 | 58.006 / 0.850 | -7.477* | -6.143* |

*Note:* "IKNet (Full)" denotes the integration of Tech and Keyword inputs. "DM Stat. (Tech)" and "DM Stat. (Keyword)" indicate the Diebold-Mariano statistics comparing IKNet with the Tech Only and Keyword Only models, respectively. Asterisks (*) denote statistical significance at the 5% level ($p < 0.05$).

consistently observed within the mid-range configurations, highlighting the critical role of balancing information content and model complexity for optimal prediction.

Table 2 summarizes the prediction performance of the proposed IKNet model and six baseline models in terms of the RMSE and SMAPE. All models were trained under the same time series regression setting. The IKNet consistently achieved the lowest RMSE and SMAPE values across most years, demonstrating superior and stable predictive performance. In particular, the best accuracy was recorded in 2024, with an RMSE of 58.006 and SMAPE of 0.850.

Although certain baseline models perform competitively in specific years, their results tend to vary significantly over time and often fail to capture fine-grained market signals. In contrast, IKNet effectively integrates both technical indicators and news-based keyword information, thereby enabling robust performance under diverse market conditions. These outcomes highlight the ability of the model to leverage keyword-level semantic signals and structured features to achieve high predictive accuracy and generalization.

To assess the contribution of news-based keyword information quantitatively, we compared three model variants based on the same Bi-LSTM architecture: one using only technical indicators (Tech only), one using only news keywords (Keywords only), and a full model (IKNet) incorporating both inputs.

Table 3 lists the RMSE and SMAPE values for each model. To determine whether the performance differences between the single-input and full models were statistically significant, we applied the Diebold-Mariano (DM) test [11], which compares the prediction error distributions for the same target points.

As shown in Table 3, in most years, both single-input models exhibited significantly higher prediction errors than did the full model. The performance gap was particularly notable in 2024 for both metrics. In addition, the Keyword only model generally outperformed the Tech only model, suggesting that news-based features may offer more direct predictive signals. However, in some years, single-input models still achieved reasonable accuracy, indicating that the relevance of each input type may vary, depending on the market conditions. These results empirically demonstrate that combining technical and news inputs effectively captures both sequential patterns and external shocks, which thereby improves the overall forecasting performance.

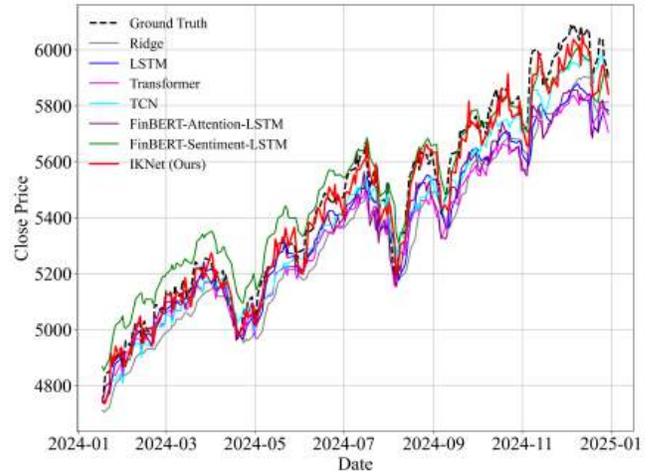

**Figure 2: Comparison of predicted and actual closing prices in 2024 for each model.**

Figure 2 presents a visual comparison between the actual closing prices (ground truth) and predictions of each model for the most recent year (2024). The IKNet closely tracks the actual price trends throughout the period and exhibits relatively small prediction errors, particularly during periods of high volatility. These results indicate that the model effectively captures complex nonlinear patterns in time series data by integrating technical indicators with financial news keyword embeddings. Moreover, IKNet responds promptly to upward trend shifts and avoids excessive deviations during downturns, suggesting robust generalization across diverse market conditions. This visual analysis complements quantitative metrics by providing qualitative insights into trends sensitivity and prediction stability.

### 5.3 Comparison of Investment Profits by Model

The return simulation is based on a trading strategy that takes a long position when an upward trend is predicted and liquidates the position when a downward trend is forecasted. A fixed transaction cost of 0.3% was applied at both entry and exit, following standard assumptions from prior studies [28]. While holding a position, log returns are computed and accumulated based on actual closing prices [6]. The log return at time step $t$, denoted by $R_t$, is defined as follows:

$$R_t = \log\left(\frac{P_t}{P_{t-1}}\right), \tag{6}$$

where $P_t$ denotes the closing price at time $t$. If the predicted direction does not match the actual price movement, or if no position is held, the return is set to zero. The cumulative return is computed by exponentiating the sum of the individual log returns and converting the result into a percentage, as follows:

$$\text{Cumulative Return} = \left(\exp\left(\sum_t R_t\right) - 1\right) \times 100. \tag{7}$$

Table 4: Cumulative profit and Sharpe ratio of IKNet and baselines for 2018–2024.

| Year | HV (%) | Metric | Long-only | Others' Best | IKNet (Ours) |
|---|---|---|---|---|---|
| 2018 | 32.41 | Cumulative profit (%) | -8.969 | -7.096 | **-3.755** |
|  |  | Sharpe ratio | -0.573 | -0.483 | -0.286 |
| 2019 | 24.18 | Cumulative profit (%) | 19.972 | 22.180 | **24.580** |
|  |  | Sharpe ratio | 1.575 | 1.893 | 2.072 |
| 2020 | 51.70 | Cumulative profit (%) | 11.257 | **35.805** | 34.651 |
|  |  | Sharpe ratio | 0.311 | 1.191 | 1.338 |
| 2021 | 40.83 | Cumulative profit (%) | **21.643** | 14.034 | 15.659 |
|  |  | Sharpe ratio | 1.537 | 1.268 | 1.525 |
| 2022 | 73.79 | Cumulative profit (%) | -13.698 | -6.203 | **-5.579** |
|  |  | Sharpe ratio | -0.640 | -0.321 | -0.304 |
| 2023 | 43.54 | Cumulative profit (%) | 18.194 | **23.269** | 21.973 |
|  |  | Sharpe ratio | 1.347 | 1.720 | 1.626 |
| 2024 | 48.81 | Cumulative profit (%) | 20.944 | 19.556 | **23.182** |
|  |  | Sharpe ratio | 1.536 | 1.481 | 1.806 |

*Note:* HV (historical volatility) represents the annualized standard deviation of the daily log returns, calculated as $HV = \sigma_{\text{daily}} \times \sqrt{N}$, where $\sigma_{\text{daily}}$ is the standard deviation of the daily log returns and $N = 252$ denotes the number of trading days in a year.

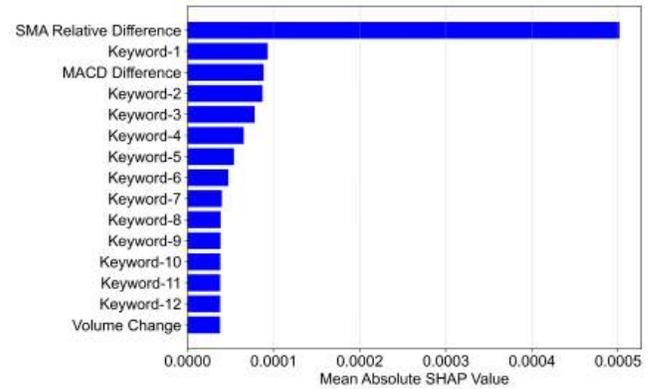

Figure 3: SHAP-based feature importance plots for 2024.

Table 4 presents a quantitative comparison of investment performance across different models based on the cumulative profit and Sharpe ratio, which are calculated on an annual basis. "Others' Best" refers to the best-performing baseline model in each year, and "Long-only" denotes a simple buy-and-hold strategy without prediction.

The IKNet consistently achieves the highest Sharpe ratio for most years, indicating its superior risk-adjusted returns. In 2020, the Sharpe ratio reached 1.338, the highest among all the models, and in 2024 it reached 1.806, demonstrating stable investment performance. Additionally, the cumulative profit of IKNet in 2024 was 23.18%, which was the highest return observed across all models.

Even in bearish markets, such as 2018 and 2022, IKNet effectively limited losses. In particular, in 2022, when the annual volatility reached 73.79%, the model showed a strong ability to avoid excessive risk exposure. These results empirically demonstrate that IKNet is not only accurate in price prediction but also practical and stable from an investment management perspective. The consistent performance of both the return and volatility metrics suggests that the model has a strong potential for real-world trading applications.

### 5.4 SHAP-based Explainability Analysis

To assess the model interpretability, we quantified the predictive contributions of technical indicators and keyword embeddings using SHAP values. Based on the 2024 test data, we visualized the average SHAP importance scores for all input features. At the article level, keyword-wise SHAP values were further visualized to evaluate the consistency and interpretability of the word-level explanations.

Figure 3 shows that news keywords consistently ranked among the top contributors, underscoring their critical influence on predictions. Some keywords even exceeded major technical indicators in importance, indicating that unstructured information functions not only as a supplementary input but also as a primary decision driver. By capturing the qualitative market context, keywords complement technical indicators by reflecting investor sentiment and shifts in market conditions that are difficult to detect.

Conversely, technical indicators are important in several cases because they offer stability and reliability as structured inputs. These results confirm the complementary roles of structured and unstructured data in the IKNet. The model effectively integrates both sources to enhance predictive accuracy, interpretability, and practical utility.

To interpret how individual keywords contributed to predictions in specific contexts, we employed SHAP-based text visualization. Figure 4 shows an example of a news article published on August 2, 2024, regarding a U.S. employment report. Each word was annotated with its SHAP value, where the sign indicates whether the word pushed the prediction upward or downward and the magnitude represents the strength of its influence. These values are visualized using arrows that encode both the direction and intensity.

The analysis revealed that negative keywords, such as "tumbled", "plunged", "layoffs", "hurt", "headwinds", and "fell" strongly contributed to lowering the predicted price. These results align with the actual market reaction, as the S&P 500 index declined by approximately 3.0% on August 5. These words reflect negative economic signals, such as deteriorating employment, weak corporate performance, and cost cutting, which collectively drive the output below the baseline. Particularly, "layoffs" and "hurt", which are directly associated with employment conditions, served as strong indicators, consistent with the tone of the report.

Conversely, positive terms such as "boosted", "expansion", and "growth" contribute to upward adjustment, but negative sentiment dominates overall, resulting in a lower final prediction. These results demonstrate that SHAP enables the fine-grained estimation of directional contributions at the keyword level. Compared with sentiment scores or document-level embeddings, the proposed approach provides more precise interpretability. Visualization offers insights into the decision process of the model to support both the explainability and reliability of news-driven stock predictions.

### 6 DISCUSSION

This study presented IKNet, which is a stock prediction model that integrates technical indicators and salient keywords from financial news sources. Its structural validity and practical effectiveness were confirmed through comparative experiments, which yielded three key findings.

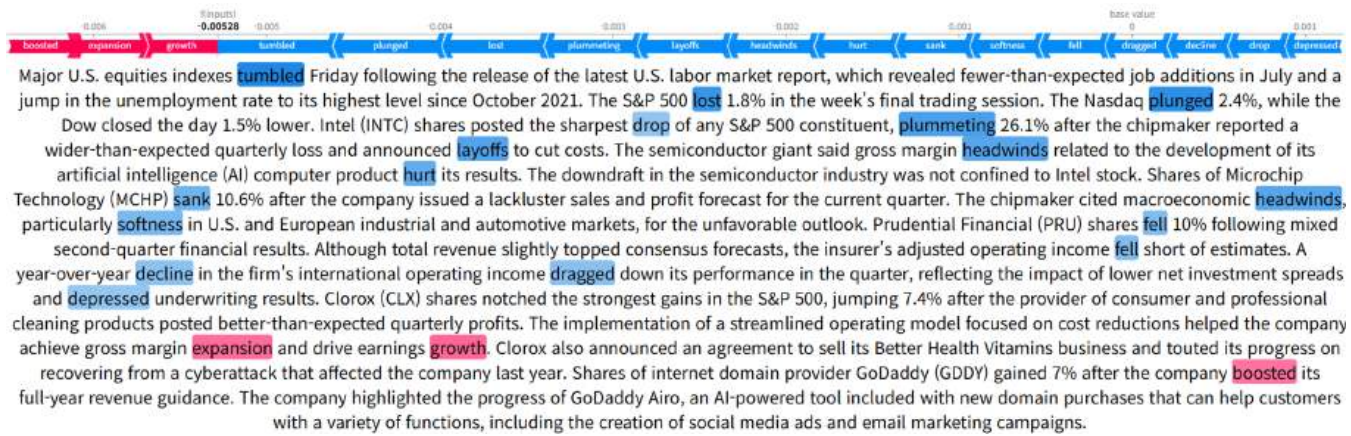

Figure 4: SHAP-based interpretation of keyword contributions in a news article published on August 2, 2024.

First, IKNet consistently outperforms traditional time-series and news-based models, and it maintains robust trend-following behavior and stable returns even under high market volatility. This performance is attributable to the selective use of core keywords from each news article, which enables the model to reflect both external events and time-series patterns more effectively.

Second, experiments on the number of input keywords showed that a moderate size minimized the RMSE and SMAPE, whereas overly large or small input sizes degraded the performance, highlighting the need to balance information richness and model complexity. Additionally, ablation studies confirmed that including salient keywords significantly enhances predictive accuracy.

Third, a SHAP-based analysis identified the most influential features behind the predictions of the model. The keyword-level SHAP visualization revealed the direction and magnitude of each contribution. In contrast to prior studies that focused on global feature importance, this study provides a fine-grained attribution of individual keywords to enable a more contextual understanding of sentiment dynamics and external narratives.

However, the current model relies on fixed keyword embeddings that may not fully capture contextual shifts in meaning. Future studies could incorporate sentence- or context-level representations to improve interpretive precision and generalizability.

## 7 CONCLUSION AND FUTURE WORK

This study proposes IKNet, which is a stock-prediction model that integrates technical indicators with keywords selectively extracted from financial news to achieve both predictive accuracy and interpretability. By adopting a keyword-level input structure and a SHAP-based explanation framework, the model addresses the limitations of conventional approaches that rely on average embeddings or sentiment scores.

The empirical results demonstrate that IKNet maintains stable performance across diverse market conditions. The complementary integration of keywords and technical features enhances forecasting accuracy, whereas SHAP-based visualizations provide clear and quantitative explanations of model outputs.

Future studies will explore encoding strategies that incorporate sentence- or context-level representations to better capture the semantic variations of keywords. In addition, dynamically adjusting the relative importance of news and technical indicators may improve the adaptability of the model to shifting market environments. Beyond numerical prediction, IKNet offers interpretable insights into qualitative news signals, making it a practical decision-support tool. The transparency and applicability of IKNet suggest strong potential for its extension to trading strategy development and broader financial services.

## Acknowledgments

This work was supported by the National Research Foundation of Korea(NRF) grant funded by the Korea government(MSIT)(No. RS-2025-00554384), and the Technology Development Program (No.RS-2024-00513926) funded by the Ministry of SMEs and Startups(MSS, Korea)